\documentclass[epj]{webofc}
\usepackage[utf8]{inputenc}
\usepackage[varg]{txfonts}   % Web of Conferences font
\usepackage{booktabs}
\usepackage{xcolor}
\definecolor{darkred}{rgb}{0.4,0.0,0.0}
\definecolor{darkgreen}{rgb}{0.0,0.4,0.0}
\definecolor{darkblue}{rgb}{0.0,0.0,0.4}
\usepackage[bookmarks,linktocpage,colorlinks,
    linkcolor = darkred,
    urlcolor  = darkblue,
    citecolor = darkgreen]{hyperref}
\usepackage{bm}
\usepackage[binary-units = true]{siunitx}
\usepackage{nicefrac}
\usepackage{listings}
\usepackage{tabularx}
\usepackage[section]{placeins}
\usepackage[strings]{underscore}
\wocname{EPJ Web of Conferences}
\woctitle{Lattice2017}
\newcommand{\wmg}{\mbox{DD-$\alpha$AMG}\xspace}
\newcommand{\ddr}{DDR4\xspace}
\newcommand{\hca}{HCA\xspace}
\newcommand{\ib}{InfiniBand\xspace}
\newcommand{\kncs}{KNCs\xspace}
\newcommand{\knc}{KNC\xspace}
\newcommand{\knls}{KNLs\xspace}
\newcommand{\knl}{KNL\xspace}
\newcommand{\lqcd}{LQCD\xspace}
\newcommand{\mcd}{MCDRAM\xspace}
\newcommand{\numa}{NUMA\xspace}
\newcommand{\opa}{Omni-Path\xspace}
\newcommand{\pmr}{pMR\xspace}
\newcommand{\psm}{PSM2\xspace}
\newcommand{\qp}[1]{QPACE~#1\xspace}
\newcommand{\vpu}{VPU\xspace}
\begin{document}
%%%%%%%%%%%%%%%%%%%%%%%%%%%%%%%%%%%%%%%%%%%%%%%%%%%%%%%%%%%%%%%%%%%%%%%%%%%%%

\selectlanguage{english}

\title{%
  \boldmath\wmg on \qp{3}
}

\author{%
\firstname{Peter} \lastname{Georg}\inst{1}\fnsep\thanks{Speaker, \email{peter.georg@ur.de}} \and
\firstname{Daniel} \lastname{Richtmann}\inst{1}\fnsep\thanks{Speaker, \email{daniel.richtmann@ur.de}} \and
\firstname{Tilo}  \lastname{Wettig}\inst{1}
}

\institute{%
Department of Physics, University of Regensburg, 93040 Regensburg, Germany
}

\abstract{%
We describe our experience porting the Regensburg implementation of the \wmg solver from \qp{2} to \qp{3}. We first review how the code was ported from the first generation Intel Xeon Phi processor (Knights Corner) to its successor (Knights Landing).
We then describe the modifications in the communication library necessitated by the switch from \ib to \opa.
Finally, we present the performance of the code on a single processor as well as the scaling on many nodes, where in both cases the speedup factor is close to the theoretical expectations.
}

\maketitle

\section{Introduction}\label{sec:intro}

The lattice QCD (\lqcd) community has traditionally been an early adopter of new computing and network architectures.
This typically requires major efforts porting simulation code or even communication libraries.
The Regensburg lattice group (RQCD) has been involved in such efforts, as well as supercomputer development, for more than a decade.
While the first computer in the QPACE series \cite{Goldrian:2008vlh, Baier:2009yq} was based on IBM's Cell processor and an FPGA-based custom interconnect, the subsequent machines are using Intel's Xeon Phi series with standard interconnects (see Sec.~\ref{subsec:overview}).
To satisfy the increasing demands of the RQCD physics program we use a state-of-the-art method, \wmg \cite{Frommer:2013fsa}, to solve the discretized form of the Dirac equation.
The high-performance implementation of this solver on \qp{2} is described in \cite{Heybrock:2014iga, Heybrock:lat15, Richtmann:2016kcq, Georg:2017diz}.
The present contribution focuses on the software efforts we made to efficiently run this implementation on \qp{3}.

This paper is structured as follows.
In Sec.~\ref{sec:qpace3} we give an overview of \qp3 and highlight the differences to \qp2 in terms of processor and network.
We discuss the network technology in some detail because it has changed rather drastically.
In Sec.~\ref{sec:portingourimplementation} we describe how our solver and our communication library were adapted to the new technologies.
In Sec.~\ref{sec:performancefigures} we present single-node and multi-node benchmarks of the solver on \qp3 and compare the results with numbers obtained on \qp2.
In Sec.~\ref{sec:conclusions} we conclude and give an outlook on future work.

\section{\qp{3}}\label{sec:qpace3}

\subsection{Overview}\label{subsec:overview}

While \qp{2} \cite{Arts:2015jia} is based on the Knights Corner (\knc) version of the Intel Xeon Phi processor series and an FDR \ib network, its successor \qp{3} utilizes the current Xeon Phi processor, Knights Landing (\knl), and Intel's new \opa fabric.
\qp{3} was installed in two phases.
Phase 1 consists of 352 nodes, each equipped with a 64-core Xeon Phi 7210 running at a clock frequency of \SI{1.3}{\giga\hertz}.
Each core can run up to 4 hardware threads, giving a total of 256 threads per chip.
The \knls have \SI{16}{\giga\byte} on-package high-bandwidth memory, denoted \mcd by Intel, as well as \SI{48}{\giga\byte} of \ddr memory.
The \opa network is arranged in a 2:1 blocking-tree topology, where we use edge switches exclusively.
This system was ranked 5th in the November 2016 issue of the Green 500 list and 18th in June 2017, both times being the most energy-efficient \knl system on the list.
Phase 2 consists of 320 nodes with almost the same configuration.
The only differences to Phase 1 are the doubled amount of \ddr memory (\SI{96}{\giga\byte} instead of \num{48}) and the blocking factor of 5:1.
Given a fixed budget, these choices were made to be able to efficiently run ensemble-generation jobs that require strong scaling to many nodes on phase 1 as well as weak-scaling analysis jobs that require more memory per node on phase 2.
All data shown in these proceedings were obtained on \qp3 (or on \qp2 when comparisons are made).

\subsection{\knc vs.\ \knl hardware comparison}\label{sec:knlhardwarefeatures}

While the \knl is quite similar to the \knc from the point of view of a software developer, with the exception that the \knl is now  self-bootable,\footnote{A PCIe card was also planned but never reached the mass market.} the hardware has been changed quite significantly.
Starting at the innermost level, the first improvement is the addition of a second vector processing unit (\vpu), which enables a \knl core to issue twice as many floating-point operations per clock.
Another improvement is that these cores are now able to execute instructions out of order, which reduces stalling penalties when cache misses occur.
Furthermore, a \knc core could issue instructions from a given thread only every other cycle, which led to the need to have at least two threads running on the same core to be able to issue instructions in each cycle.
This restriction is no longer present on the \knl.
Concerning the cache structure,
the per-core size of the L1 and L2 caches stays constant at \SI{32}{\kilo\byte} and \SI{512}{\kilo\byte}, respectively.
However, with the \knl, Intel introduces the concept of a tile, which bundles two cores together sharing \SI{1}{\mega\byte} of L2 cache.
In contrast to standard Xeon processors, there is no shared L3 cache.
However, Intel makes up for that by adding an on-package high-bandwidth memory (\mcd) with a capacity of \SI{16}{\giga\byte} and a bandwidth of about \SI{420}{\giga\byte\per\second}, in addition to a standard \ddr memory interface  with 6 channels and a bandwidth of about \SI{80}{\giga\byte\per\second} (both numbers are for the \knl 7210, to be compared with about \SI{160}{\giga\byte\per\second} for the \knc 7120).
The \mcd is probably the most significant new feature of the \knl.
There are different usage models for it, called memory modes.
It can either be used as a large L3 cache (Cache mode), as a directly mapped \numa node yielding an extra \SI{16}{\giga\byte} of memory in addition to the \ddr memory (Flat mode), or a combination of both, called Hybrid mode.
The tiles, the \mcd and \ddr memory controllers, and the distributed tag directory are connected in a two-dimensional mesh, in contrast to the \knc's ring bus.
According to Intel, this yields higher bandwidth and lower latency between the cores.
This feature is by now also incorporated in the Xeon server architecture.
The 2D mesh increases the flexibility of configuration options of the \knl even further with so-called cluster modes.
It enables the chip to be used either as is (All-to-All) or divided into two/four equal parts which are then either software transparent (Hemisphere/Quadrant) or exposed to the operating system as separate \numa domains (SNC2/4).
From first to last, these modes increase the affinity between tile, distributed tag directory, and memory, thus yielding lower latency and higher bandwidth.

\subsection{\ib vs.\ \opa hardware comparison}\label{subsec:opavsib}

\qp3 utilizes the \opa interconnect, replacing the previously used \ib, for communication in multi-node jobs and access to the shared network storage.
As shown in previous work \cite{Georg:2017diz} optimization for a particular network may yield significant improvements compared to relying on plain MPI.
Therefore our \wmg implementation uses a custom communication library, \pmr.
Previously, \pmr only supported \ib and local inter-process communication, leveraging Linux CMA.
The idea now is to add support for \opa to \pmr. 
To do that properly, it is crucial to understand the hardware, especially the differences to the well-known \ib hardware.
Apart from many differences that do not affect us, there are two main differences of these competing technologies we need to take a closer look at: connection-oriented vs.\ connectionless communication and interconnect offloading vs.\ onloading.

\subsubsection{Connection-oriented vs.\ connectionless communication }\label{subsubsec:connection-oriented-vs-connectionless-communication}

\opa implements connectionless communication only, while \ib mainly relies on connection-oriented communication for reliable data transfer.
The \ib specification also includes connectionless reliable and unreliable data transfers, but the former is not implemented in any well-known hardware.
For simplicity, we ignore unreliable data transfer as it is not used in any communication pattern of interest to us.

Connection-oriented communication requires one connection for each pair of processes that communicate with each other.
Each connection consumes a certain amount of host memory, and the total memory utilization scales linearly with the number of connections.
The \ib Host Channel Adapter (\hca) uses on-chip memory to cache connection-related data,
but if the cache is full it has to exchange data with the host's main memory via PCIe, resulting in a performance penalty.
It is therefore sensible to minimize the number of connections.

In a connectionless approach it is only necessary to set up communication endpoints and exchange address information with all other peers once (with MPI, typically during the initialization phase).
This allows for scaling of applications to an arbitrary number of processes without any noticeable increase in the resources required per process.

The preceding discussion implies that connectionless communication should be superior. 
For communication patterns that rely heavily on all-to-all communication, this indeed seems to be the case \cite{Mamidala:2007:UCV:1229428.1229437}.
However, there are two reasons why connection-oriented communication can still obtain similar or better performance.

First, many stencil-type applications, including \lqcd, only require a limited subset of communication patterns for performance-relevant parts, in particular nearest-neighbor halo exchanges and global reductions.
The former requires a maximum of $2d$ connections per process in a $d$-dimensional theory.
The latter is often implemented using the recursive-doubling algorithm, which in turn consists of nearest-neighbor exchanges in a $\log_2(p)$-dimensional grid, where $p$ is the number of processes.
Hence, it only adds $\log_2(p)$ connections, some of which might even be identical to the connections already set up for halo exchanges.
Assuming a 256-node job with one process per node, no more than $8 + \log_2(256) = 16 $ connections are required to be set up for those two communication patterns.
For other non performance-relevant parts, e.g., parallel input/output to the shared network storage, connections can be set up dynamically without any noticeable impact on wall-clock time. 

Second, a number of hard- and software features have been developed to reduce the number of connections and thus alleviate the main drawback of connection-oriented communication.
In software, many MPI implementations allow for changing from a static to a dynamic connection setup.
In hardware, new connection modes have been added, such as the Extended Reliable Connected (XRC) Transport Service that enables processes running on the same node to share certain connections \cite{koop:2008} and the Dynamically Connected (DC) Transport service that hands over the connection management to the \hca, which then sets up or tears down connections dynamically as required \cite{Subramoni:2014:DML:2769884.2769903}.
However, if the number of connections is low, as in stencil-type applications, these new modes are not necessary and the traditional reliable connected (RC) mode still yields best performance.

\subsubsection{Interconnect offloading vs.\ onloading}\label{interconnect-offloading-vs-onloading}

Interconnect offloading/onloading specifies whether network functions are offloaded to the network adapter (\ib \hca), or onloaded onto the CPU (\opa).

Onloading interconnect technology tends to be less complex and hence easier to build.
However, while the CPU is managing and executing network operations it is not available for other tasks, most importantly computations. 
Whether this CPU overhead is significant depends on the particular application. 

Offloading hardware does not block any compute resources and can be beneficial in two ways.
First, if an application is able to overlap communication and computation, the CPU can continue to execute computation tasks while the network adapter performs communication.
This is true for \lqcd simulations, where many algorithms allow for overlap of computation and halo exchanges.
Second, even if no overlap is possible, offloading enables several technologies that can improve performance.
One example is RDMA, which reduces the latency involved in non-overlapping halo exchanges.
Another example is relevant for the other important communication pattern in \lqcd, i.e., global reductions.
These can hardly be overlapped with computation, as the result is either required immediately or is used as a stopping criterion in iterative algorithms.
Most \ib adapters nowadays support offloading collective routines to the \hca, reducing the involvement of the host CPU in these operations \cite{Graham:2010:CIM:1844765.1845221}.
For global reductions, this approach is taken even further with the most recent \ib networks, which now support in-network computing and in-network memory to reduce data movement \cite{Graham:2016:SHA:3018058.3018059}.

\section{Porting of simulation code and communication library}\label{sec:portingourimplementation}

\subsection{\boldmath\wmg for Xeon Phi}

The starting point of the present work is our implementation \cite{Heybrock:lat15,Richtmann:2016kcq} of the \wmg algorithm for \qp{2}.
This implementation contains a number of optimizations with respect to the original Wuppertal code (which was neither threaded nor vectorized).
In the following we describe these optimizations, most of which are now part of the official \wmg code base \cite{ddalphaamggithubrepo}.
We adapted the data layout to be able to make efficient use of the hardware (i.e., the caches and the \SI{512}{\bit} vector registers), performed extensive vectorization using compiler intrinsics, and inserted many software prefetching directives to overcome the limitations of the \knc.
Furthermore, we implemented an MPI-like threading model with fully persistent threads and further enhanced the use of mixed precision by adding support for storing some data structures in half precision, i.e., \SI{16}{\bit} floating-point numbers.

The subject of this contribution is the porting of our existing code base to \qp{3}.
Reference~\cite{Heybrock:2014iga}, which was prepared in collaboration with Intel engineers and is part of our code base as the fine-grid smoother, states that the port from \knc to \knl will require only modest efforts since the instruction set architecture of these two processors is quite similar.
Mainly, we need to replace the explicit IMCI intrinsics scattered all around the code by the corresponding AVX512 intrinsics.
We realize this by introducing a lightweight abstraction layer consisting of small functions.
This layer wraps the bare intrinsics, which is mostly straightforward except for the permutation intrinsics.
At the time of this writing, no further \knl-specific optimizations have been performed.

An issue of particular interest for our porting efforts is the support for half precision.
The \knc does not support computations in half precision, but is able to do on-the-fly up/down conversions in hardware when loading/storing data from/to memory.
We utilize this feature in our code to reduce the working set size, and as a consequence the requirements on cache capacity and memory bandwidth.
However, on the \knl the up/down conversions are no longer part of the instruction set.
We attempted to work around this problem by utilizing legacy Intel processor instructions, as depicted in Listing~\ref{lst:half-prec-masked-store}.
Unfortunately, this attempt actually degrades the performance, see the benchmarks below.

\begin{lstlisting}[float,caption={Wrapper function for a masked store operation from a SIMD register to memory in half precision. The \knl code is inspired by the up/down conversion in the code generator of the QPhiX library \cite{qphixgithubrepo}.},captionpos=b,label={lst:half-prec-masked-store}]
inline void store(half *data, maskF mask, regF &reg)
{
#if defined(KNC)
    _mm512_mask_extstore_ps(data, mask, reg, _MM_DOWNCONV_PS_FLOAT16, 0);

#elif defined(KNL)
    _mm256_store_si256(
        (__m256i *)data,
        _mm512_cvtps_ph(
            _mm512_mask_blend_ps(
                mask,
                _mm512_cvtph_ps(_mm256_load_si256((__m256i const *)data)),
                reg),
            _MM_FROUND_TO_NEAREST_INT));

#endif
}
\end{lstlisting}

In contrast to the \knc, where only the Intel compiler and its runtime libraries are usable, the \knl is supported by all three major compiler suites: GCC, Clang, and Intel.
In Tables~\ref{tab:comparison-single-core-numbers-mr-inversion} and \ref{tab:comparison-single-core-numbers-schwarz-method} we compare the single-core performance of two important code parts, the MR inversion within a domain and the entire Schwarz method, obtained with these compilers.\footnote{Using GCC we were unable to compile the code with half-precision support turned on.}

\begin{table}[thb]
    \begin{center}
        \begin{tabular}{clrr}
            Processor & Compiler     & GFlop/s single & GFlop/s half \\
            \midrule
            \knc & Intel 15.0.3 & 8.9           & 10.3        \\
            \knl & GCC 6.2.1    & 8.5           & X           \\
            \knl & Clang 4.0.0  & 13.8          & 11.3        \\
            \knl & Intel 17.0.2 & 16.6          & 11.4        \\[-5mm]
        \end{tabular}
    \end{center}
    \caption{Single-core performance for the MR inversion within a domain.\vspace*{-5mm}}
    \label{tab:comparison-single-core-numbers-mr-inversion}
\end{table}
\begin{table}[thb]
    \begin{center}
        \begin{tabular}{clrr}
            Processor & Compiler     & GFlop/s single & GFlop/s half \\
            \midrule
            \knc & Intel 15.0.3 & 7.4           & 9.0         \\
            \knl & GCC 6.2.1    & 5.6           & X           \\
            \knl & Clang 4.0.0  & 9.2           & 7.6         \\
            \knl & Intel 17.0.2 & 9.7           & 7.6         \\[-5mm]
        \end{tabular}
    \end{center}
    \caption{Single-core performance for the Schwarz method.}
    \label{tab:comparison-single-core-numbers-schwarz-method}
\end{table}

\noindent 
We see that the Intel compiler yields the best performance, closely followed by Clang.
On \knl, half precision deteriorates the performance, rather than improving it, in contrast to \knc.
This may be a problem with our implementation and will be investigated further in the future.

The \knc has no L1 hardware prefetcher, and the performance of the L2 prefetcher is not optimal.
Therefore software prefetching of data to the caches is essential for achieving good performance on \knc.
Although the \knl now features an L1 hardware prefetcher, we still use manual software prefetching since our code base already contains these directives.

\subsection{\pmr communication library}\label{subsec:opasoftware}

In Sec.~\ref{subsec:opavsib} we have discussed the hardware differences of \ib and \opa.
In this section we take a look at the \opa software stack to identify necessary modifications of our own communication library.

There are currently two APIs available to work with \opa.
The first API, which MPI implementors have been encouraged to rely on, is defined by libfabric, which is a core component of OpenFabrics Interfaces (OFI), a common framework for various interconnects.
This API is supposed to be hardware agnostic and yield good performance on any supported hardware.
However, this is currently not the case.
Although one can, in theory, use the same user code on top of this API on, e.g., \ib and \opa hardware, this leads to severe performance degradation.
Therefore, in practice the user code needs to be adapted to the hardware to achieve good performance.
The second API is Performance Scaled Messaging 2 (\psm), which is used by libfabric under the hood for \opa.
Due to the current limitations of libfabric just described, it is sensible to use \psm directly to avoid performance degradation due to unsuitable abstraction.
This is in fact what many MPI implementations still do, and we also follow this path for \pmr.

The \psm API supports a tag-matched two-sided messaging model that is very similar to the MPI two-sided messaging API.
In contrast to \ib it is not necessary to register any memory region for use as send or receive buffers.
Although \psm is the lowest-level \opa API, it not only supports \opa but also includes a self (for intra-process communication) and a shared-memory (for local inter-process communication) provider.
Furthermore, \psm over \opa supports two send methods, programmed input/output (PIO) and Direct Memory Access (DMA), and two receive methods, Token ID (TID, commonly known as rendezvous) and eager.
These methods are mainly chosen by globally set thresholds and can hardly be influenced by the user code.
The send method has an impact on the CPU utilization.
As for the receive methods, the eager protocol might be able to reduce network latency by adding an additional memory copy on the receiving side.

\begin{figure}[t]
    \begin{center}
        \includegraphics{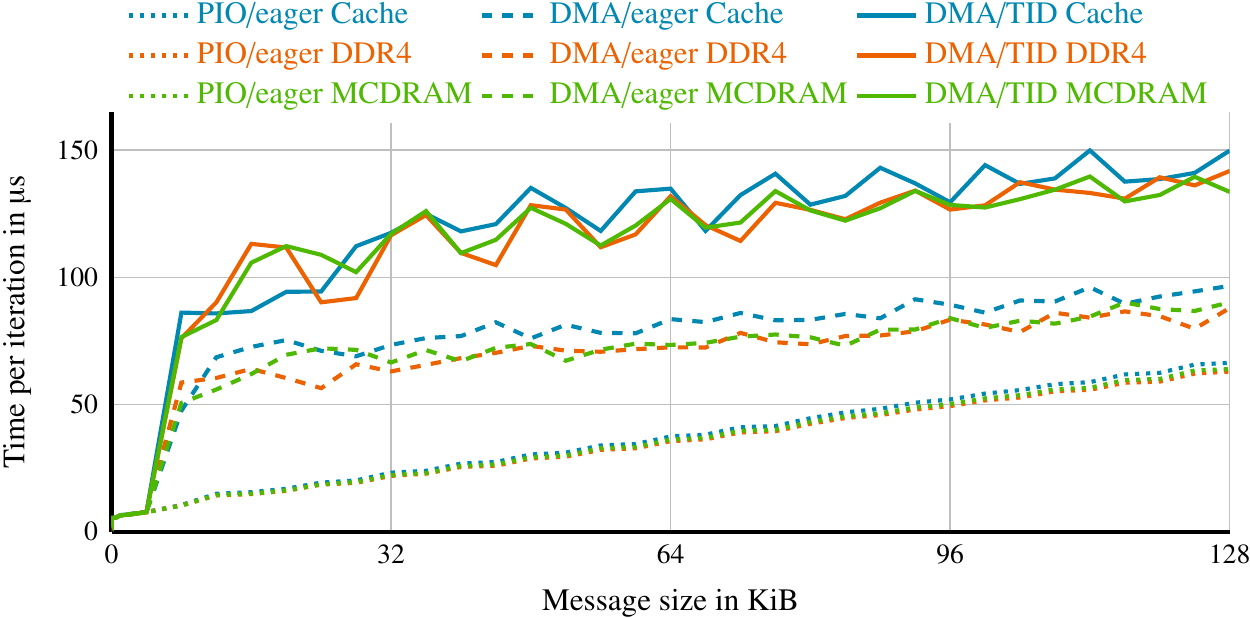}
    \end{center}
    \caption{Ping-ping benchmark of \psm send and receive methods as a function of the \knl memory modes, cache and flat. For flat mode the application was bound once to \ddr and once to \mcd. The combination PIO/TID is not available.}
    \label{fig:hfimethods}
\end{figure}

In Fig.~\ref{fig:hfimethods} we compare the performance of all available combinations of \psm send and receive methods as a function of the \knl memory modes.
For all three combinations there are only marginal differences between \mcd and \ddr, which presumably are due to the slightly lower latency of \ddr.
For PIO/eager and DMA/TID there is no significant dependence on the memory mode, while for DMA/eager the performance of the additional memory copy depends on the memory mode, with cache being worse than flat.  
Note that this synthetic benchmark is only useful to identify the best memory mode.
It cannot identify the best combination of send/receive methods because it neglects the CPU utilization.
The global thresholds at which \psm switches between different methods should be tuned using benchmarks of the actual application. 

\begin{figure}[t]
    \begin{center}
        \includegraphics{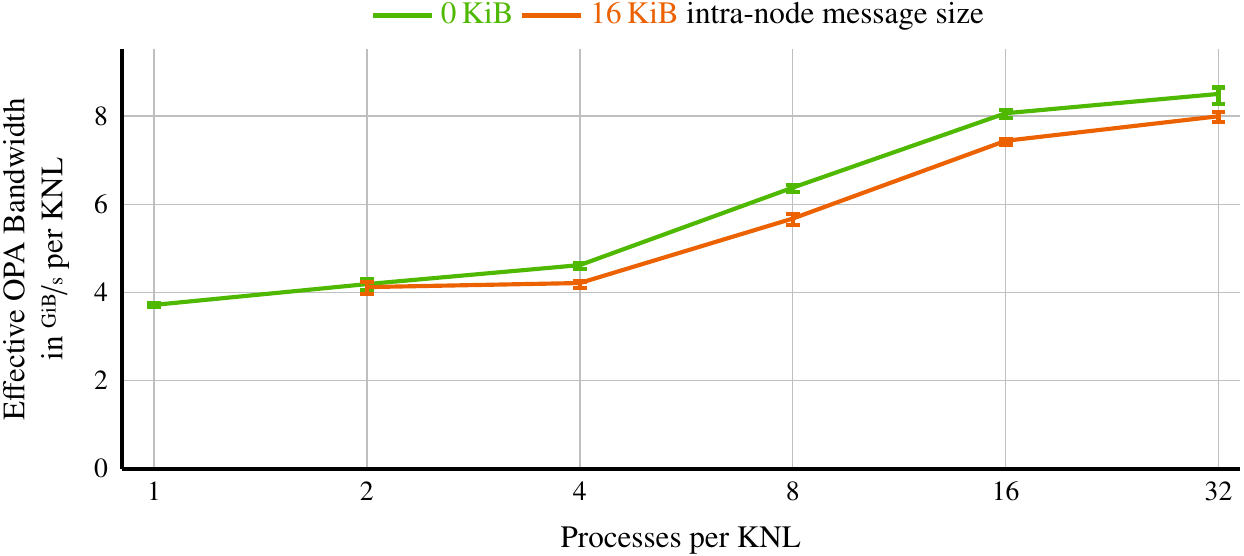}
    \end{center}
    \caption{Halo-exchange benchmark running on 16 \knl nodes. Inter-node message size per process was set to \SI{512}{\kibi\byte} / (number of processes per node) to have a constant amount of data per node transferred via \opa. For the \SI{0}{\kibi\byte} intra-node message size, the processes were synchronized to ensure that all processes are communicating at the same time and not sequentially. The \SI{16}{\kibi\byte} intra-node message size was chosen to simulate real-world applications.}
    \label{fig:hfiprocs}
\end{figure}

Since the \psm API is very similar to MPI it is easy to port existing MPI software to run directly on top of \psm.
However, there is a limitation in \psm which potentially impacts performance: \psm is limited to open only one endpoint per process,\footnote{After completion of the work presented here, which is based on  \opa Fabric Software 10.3, this limitation has been removed in version 10.5.  See Sec.~\ref{sec:conclusions} for future opportunities utilizing the updated software.} but one endpoint is not sufficient to achieve full bandwidth utilization (see Fig.~\ref{fig:hfiprocs} below). 
For non-threaded MPI applications this is obviously not an issue since there are many processes, and therefore many endpoints, per node. However, all of our software is based on hybrid MPI + OpenMP \cite{Heybrock:2014iga,Heybrock:lat15}.
To obtain full bandwidth we need to open more than one endpoint, and thus we need more than one process per node.
This in turn adds inter-process communication overhead within every node.
This intra-node overhead is not negligible but can be reduced by utilizing fully threaded communication.

In  Fig.~\ref{fig:hfiprocs} we show the effective \opa bandwidth per \knl depending on the number of processes per \knl.
The effective bandwidth increases with the number of processes (= endpoints), and we need at least 16 processes per \knl to get close to peak bandwidth.

Apart from the performance issues just discussed, implementation issues arise as well.
We want to use \psm directly for data transfer in performance-critical parts but still rely on MPI for all other parts.
Because of the limitation to one endpoint per process, either our communication library \pmr or the MPI implementation can open the endpoint, but not both.
The other party has to look up the existing endpoint, and access to the endpoint has to be managed.
To circumvent this issue we have introduced a new library between \psm and the upper layers (i.e., \pmr and MPI) which is responsible for opening the endpoint and managing access to it.
This library is injected using the preloading feature of the dynamic linker.

\section{Performance figures for \boldmath \wmg}\label{sec:performancefigures}

\subsection{On-chip strong scaling}\label{subsec:onchipstrongscaling}

We first investigate the on-chip scaling behavior of \wmg, i.e., the scaling of the wall-clock time for a single solve with the number of cores utilized on a single Xeon Phi.
Both on \knc and \knl we use the standard hybrid approach with one process per chip and threads on all cores (in our case, 4 threads per core).
All results are for a $16^3\times32$ lattice, which fits in the \SI{16}{\giga\byte} corresponding to the total memory of a \knc or the \mcd of a \knl.
Before discussing the results we remind the reader of the relevant peak performance figures.
The ratio of the peak floating-point performance of \knl and \knc is 2.2.
The memory bandwidth is about \SI{420}{\giga\byte\per\second} for \knl with \mcd, \SI{80}{\giga\byte\per\second} for \knl with \ddr, and \SI{160}{\giga\byte\per\second} for \knc, respectively, i.e., the ratio of \knl with \mcd to \knc is about 2.6.

\begin{figure}[thb]
    \begin{center}
        \includegraphics{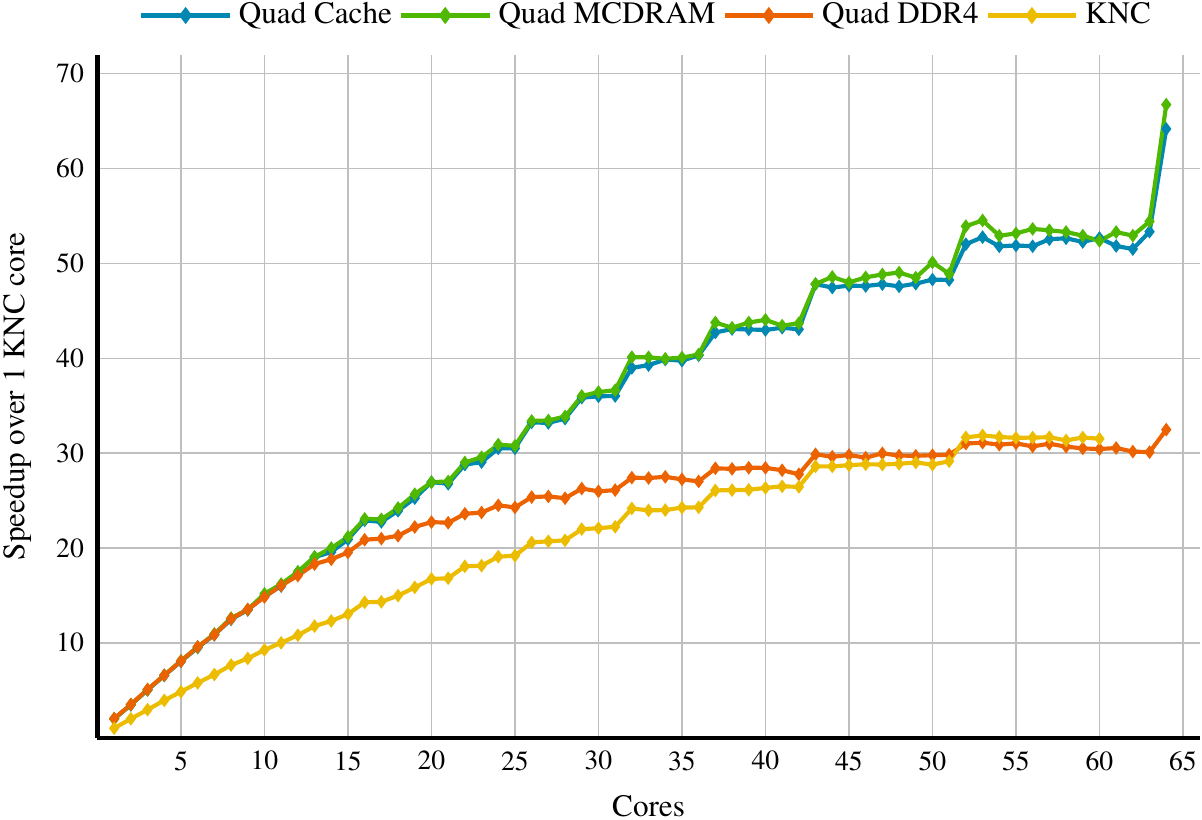}
    \end{center}
    \caption{On-chip strong scaling of the \wmg solver for a $16^3\times32$ lattice on \knc and on \knl in different memory modes.}
    \label{fig:on-chip-scaling-speedup-vs-cores}
\end{figure}

Our results are depicted in Fig.~\ref{fig:on-chip-scaling-speedup-vs-cores}, where all numbers are normalized to the value of a single \knc core.
We first notice that on the \knl there is no significant difference between cache mode and flat mode from \mcd, and therefore we do not consider cache mode further.
The remaining results should be interpreted with care.
Note that we are benchmarking a complex code and not just a simple kernel.
The \wmg code contains many parts, some of which are memory-bandwidth bound, while others are compute bound. 
Microbenchmarks have shown that the total memory bandwidth of both \knc and \knl scales roughly with the number of cores utilized. (For \knl with \ddr this statement only holds for low core count.)
Therefore, for both memory-bandwidth and compute bound code, the ideal scaling would be linear in the number of cores. 
However, once more and more cores are utilized, the overhead for synchronization between threads leads to a flattening of the speedup curve.
This is indeed what we observe qualitatively in all cases.

The case of \knl with \ddr deserves special attention.
For low core count the memory bandwidth per core is the same for \mcd and \ddr, and therefore the performance is the same in this case. 
Since both the compute power and the memory bandwidth per core is higher than on the \knc, the performance is also higher than on the \knc.
At $10\sim15$ cores we are starting to approach the sustainable \ddr bandwidth of about \SI{80}{\giga\byte\per\second} (measured with the STREAM benchmark), which explains that the red curve flattens much earlier than the green curve.
At maximum core count, \knl with DDR4 and \knc achieve about the same performance.
We regard this as an interesting coincidence.
For our particular code, it seems that the two competing effects of higher compute power and lower memory bandwidth, which affect different code parts in different ways, just compensate each other in terms of the total solve time.

Finally, the maximum performance on \knl with \mcd is a factor of 2.1 higher than on \knc, which is roughly consistent with the factors of 2.2 or 2.6 based on peak performance and memory bandwidth, respectively.

\subsection{Multi-node benchmarks}\label{sec:solverruntimemultinode}

\begin{figure}[t]
    \begin{center}
        \includegraphics{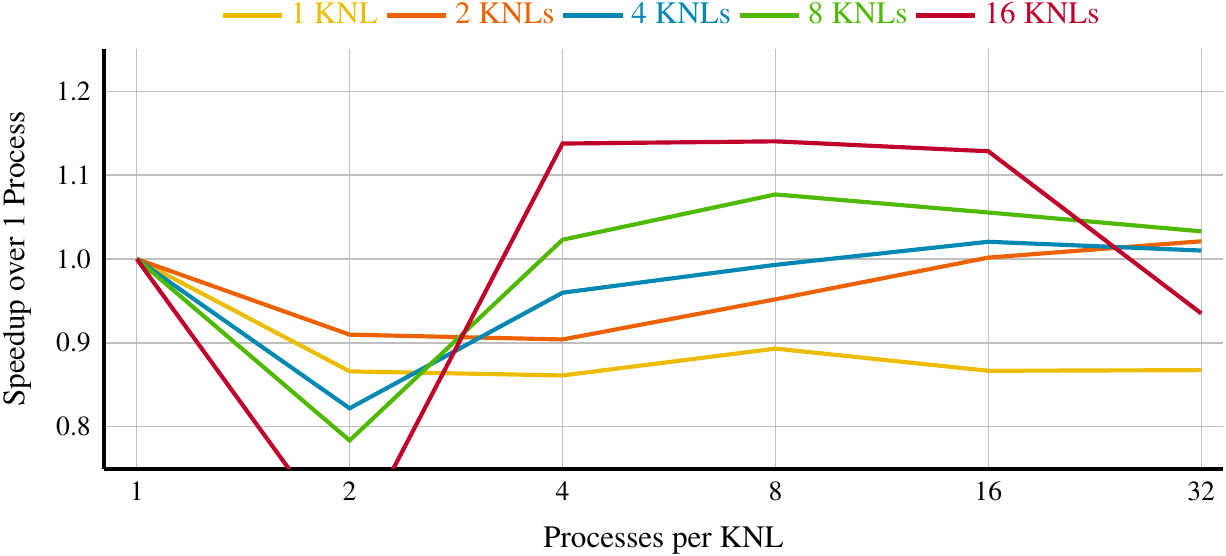}
    \end{center}
    \caption{\wmg solver with \opa on various numbers of \knls vs.\ number of processes per \knl. The lattices sizes are $16^3\times32$ for single node and $32^3\times96$ for multiple nodes.}
    \label{fig:wmgprocs}
\end{figure}

Having evaluated \opa using synthetic benchmarks, we now move to real-world benchmarks for our \wmg implementation.
In Fig.~\ref{fig:wmgprocs} we study the dependence of the performance on the number of processes per \knl node.
Depending on the number of processes per node different cluster modes are chosen: quadrant mode for 
a single process, SNC2 for two processes, and SNC4 for four or more processes.
The results can be summarized as follows: single-node performance is best with a single process, while multi-node performance can benefit from using four or more processes per node.
The former is as expected, and the latter is consistent with the synthetic benchmark in Fig.~\ref{fig:hfiprocs}.
As explained in Sec.~\ref{subsec:opasoftware}, there are two competing effects: using more processes per \knl makes more endpoints available and thus increases bandwidth utilization, but the resulting intra-node overhead reduces this effect for a low number of \knls.

\begin{figure}[thb]
    \begin{center}
        \includegraphics{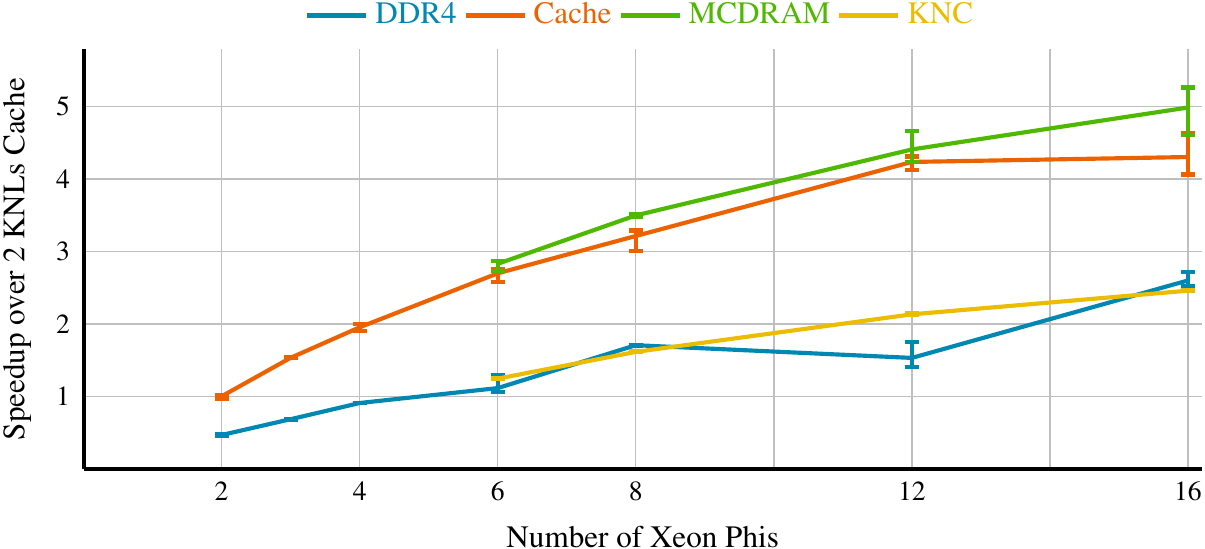}
    \end{center}
    \caption{Off-chip strong scaling of \wmg for a $32^3\times96$ lattice on \knl with \opa and on \knc with \ib. The following parameters have been tuned to achieve the best performance: number of processes per \knl, \opa threshold values (see Sec.~\ref{subsec:opasoftware}), and mapping of lattice to processes. The parameters of the solver algorithm are identical for all runs.}
    \label{fig:wmgstrongscaling}
\end{figure}

In the second benchmark, see Fig.~\ref{fig:wmgstrongscaling}, we plot the off-chip strong scaling of our \wmg implementation, comparing \knl and \opa to \knc and \ib. 
The \knc benchmarks were run on \qp2, where each \knc shares its dual-port FDR \ib adapter with three other \kncs.
Thus, the network bandwidth per \knc is limited to \SI{28}{\giga\bit\per\second} compared to \SI{100}{\giga\bit\per\second} in case of \knl.
We first observe that \knl with \ddr gives almost the same performance as \knc, which is consistent with Fig.~\ref{fig:on-chip-scaling-speedup-vs-cores}, where the single-node performance is also the same.
This means that the network bandwidth is not the limiting factor for our particular application, since quadrupling the network bandwidth from \ib to \opa does not improve the scaling behavior.
The real potential of the \knl can be unleashed by utilizing the \mcd either as cache or exclusively in flat mode.
The \opa performance is slightly worse using cache mode than running exclusively from \mcd, as already indicated by the benchmark in Fig.~\ref{fig:hfimethods}.
Running in flat mode we indeed achieve the expected speedup of about 2.1 over \knc.

\begin{figure}[thb]
    \begin{center}
        \includegraphics{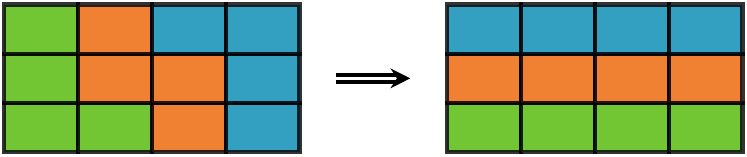}
    \end{center}
    \caption{Depiction of mapping of lattice to processes with multiple processes per node. Each rectangle presents data on one process. Rectangles of the same color represent processes running on the same node. Left: non-optimal default distribution, right: optimal distribution.}
    \label{fig:distribution}
\end{figure}

So far we have not discussed the performance drops at 6 and 12 \knls with \ddr.
These drops can be explained by a non-optimal mapping of the lattice to processes (i.e., MPI ranks) as done by QDP++ \cite{Edwards:2004sx},
which is hard to explain in words but easily shown in a picture, see Fig.~\ref{fig:distribution}.
This can only happen if there is more than one process per node and if the number of nodes contains a prime factor (three in our case) that is not contained in the number of processes per node.
The problem can easily be fixed by changing the order in which the lattice is distributed to processes, see Fig.~\ref{fig:distribution} again.
This would have to be done either in the QDP++ library or by instructing the process manager to do so.\footnote{Unfortunately, Intel MPI 2017, which was used to perform the benchmarks, contains a bug that prevents us from reversing the order in which MPI ranks are distributed to nodes.} 
An additional reason for the performance drop is that our communication is not yet fully threaded, e.g., it can happen that an inter-node data transfer is stalled by a blocking intra-node data transfer.
This is only the case when using \pmr, as the user is responsible for all threading.
In case of MPI, even if communication is not threaded in an application, the MPI implementation may have some internal threading.

\section{Conclusions and future opportunities}\label{sec:conclusions}

The subject of this contribution was the port of our existing code base for \qp{2} to our new machine \qp{3}.
We performed a minimal-efforts port of the \wmg solver by adapting the code base to the \knl instruction set architecture and retaining already existing optimizations, but not implementing any new \knl-specific optimizations.
On \knc we could achieve a significant performance gain using half precision, but on \knl 
half precision deteriorates performance rather than improving it, at least with our current implementation.
Trying out different compilers, which was not possible on \knc, we found that the Intel compiler yields best performance but has fair competition by Clang, which is thus a valuable open-source alternative.
In terms of interconnect hardware, we found that for \wmg the network bandwidth is not a bottleneck.
The key performance factors are network latency and message rate.
When running on a single or a small number of \knls, we found that the standard hybrid parallelization approach with one MPI rank per processor and threads on each core still gives the best performance for our code.
However, when running on many nodes, the best performance was obtained using multiple MPI ranks per processor, due to limitations of the \opa software stack used in this work.
The total speedup factor going from \qp2 to \qp3 is about 2.1, which we only reach when running from \mcd.

Our future strategy thus includes optimizing for flat mode, which means that the entire solver, which is an external library used by Chroma \cite{Edwards:2004sx}, is allocated in \mcd by default.
The corresponding data copying is not an issue at all, since data layout transformations are required at solver entry anyway.
The limitation on the number of endpoints per process imposed by the \opa software stack has been removed in a recent major software update.
Hence, we are no longer required to have more than one process per \knl to achieve full \opa performance but can use threaded communication and have each communication thread open its own endpoint.
Finally, we plan to improve our multi-node scaling behavior by applying domain decomposition to the coarsest grid of the multigrid method, a project that has already been started but is still work in progress.

\bibliography{lattice2017}

%%%%%%%%%%%%%%%%%%%%%%%%%%%%%%%%%%%%%%%%%%%%%%%%%%%%%%%%%%%%%%%%%%%%%%%%%%%%%
\end{document}